\documentclass[aps,pre,preprint,showpacs]{revtex4}

\usepackage{graphicx}
\usepackage{amsmath}
\usepackage{amssymb}
\usepackage{enumerate}

\DeclareMathAlphabet{\bi}{OML}{cmm}{b}{it}
\newcommand{\vx}{{\bi x}}
\newcommand{\vy}{{\bi y}}
\newcommand{\vq}{{\bi q}}
\newcommand{\vs}{{\bi s}}

\begin{document}

\title{First-principles derivation of density functional formalism for
quenched-annealed systems}

\author{Luis Lafuente}
\email{llafuent@math.uc3m.es}
\author{Jos\'e A.~Cuesta}
\email{cuesta@math.uc3m.es}
\affiliation{Grupo Interdisciplinar de Sistemas Complejos (GISC),
Departamento de Matem\'aticas, Universidad Carlos III de Madrid,
Avenida de la Universidad 30, E--28911 Legan\'es, Madrid, Spain}

\begin{abstract}
We derive from first principles (without resorting to the replica
trick) a density functional theory for fluids in quenched disordered
matrices (QA-DFT). We show that the disorder-averaged free energy of the fluid
is a functional of the average density profile of the fluid as well
as the pair correlation of the fluid and matrix particles. For practical
reasons it is preferable to use another functional: the disorder-averaged
free energy plus the fluid-matrix interaction energy, which, for fixed
fluid-matrix interaction potential, is a functional only of the average
density profile of the fluid. When the matrix is created as a quenched
configuration of another fluid, the functional can be regarded as
depending on the density profile of the matrix fluid as well. In this
situation, the replica-Ornstein-Zernike equations which do not contain
the blocking parts of the correlations can be obtained as functional
identities in this formalism, provided the second derivative of this functional
is interpreted as the \emph{connected} part of the direct correlation
function. The blocking correlations are totally absent
from QA-DFT, but nevertheless the thermodynamics can be entirely obtained
from the functional. We apply the formalism to obtain the exact functional
for an ideal fluid in an arbitrary matrix, and discuss possible
approximations for non-ideal fluids.
\end{abstract}


\maketitle

\section{Introduction}

The phase behaviour of fluids in quenched disordered matrices
has been of prior interest in the last decade. The classical
theoretical approach to these systems amounts to consider two different sets
of state variables: the \textsl{annealed} variables (usually the position of the
fluid particles), which are allowed to equilibrate,
and the \textsl{quenched} variables (usually the position of the matrix particles),
which have their values fixed. The reason for this distinction
is that our system is not in thermal equilibrium with respect to the
quenched variables, but it is in equilibrium with respect to the
annealed variables for each fixed configuration of the quenched ones.
Accordingly, two different statistical averages must be considered: the annealed average,
which is the typical ensemble average of equilibrium systems, and the quenched average
or average over disorder, which is performed over the quenched variables.
For each configuration of the disorder, we can compute the equilibrium thermodynamic
magnitudes of the system by means of the corresponding annealed
averages. These averages will, of course, depend on the configuration of the
quenched variables. However, if the matrix is statistically homogeneous (its
statistical features are similar everywhere) and the system is large, we expect
little variation between annealed averages corresponding to different matrix
configurations. Thus, quenched averages of the annealed averages are
meaningful to characterize thermodynamic magnitudes of these systems.
Because of this double average, the problem becomes intractable within the
classical equilibrium statistical-mechanics tools and new theoretical methods
are called for.

Madden and Glandt~\cite{madden:1988} made an extension of the conventional
diagrammatic treatment of liquid-state theory to obtain cluster expansions
for the thermodynamics and structure of a fluid in a quenched matrix.
They also derived a set of Ornstein-Zernike (OZ) equations relating the
total and direct interparticle correlation functions, which can be solved
with the appropriate closure relations~\cite{fanti:1990}. Alternatively,
Given and Stell~\cite{given:1992a} used the continuum version of the
replica trick~\cite{given:1992b} to rederive this set of OZ equations and
they noted that, although Madden and Glandt's cluster
expansions were correct, there were some missing terms in the OZ equations.
The corrected set of OZ equations was called the \textsl{replica} Ornstein-Zernike
(ROZ) equations and, for fluids with quenched-averaged density profile $\rho_1(\vx)$ and
matrices with one-particle distribution $\rho_0(\vx)$, is given by
\begin{align}
%h_{00}(\vx_1,\vx_2)&=c_{00}(\vx_1,\vx_2)+
%\int\mathrm{d}\vx_3 c_{00}(\vx_1,\vx_3)\rho_0(\vx_3)h_{00}(\vx_3,\vx_2),\\
%h_{10}(\vx_1,\vx_2)&=c_{10}(\vx_1,\vx_2)+
%\int\mathrm{d}\vx_3 c_{10}(\vx_1,\vx_3)\rho_0(\vx_3)h_{00}(\vx_3,\vx_2)
%+\int \mathrm{d}\vx_3 c_\mathrm{c}(\vx_1,\vx_3)\rho_1(\vx_3)h_{10}(\vx_3,\vx_2),\\
%h_{01}(\vx_1,\vx_2)&=c_{01}(\vx_1,\vx_2)+
%\int\mathrm{d}\vx_3 c_{00}(\vx_1,\vx_3)\rho_0(\vx_3)h_{01}(\vx_3,\vx_2)
%+\int \mathrm{d}\vx_3 c_{01}(\vx_1,\vx_3)\rho_1(\vx_3)h_\mathrm{c}(\vx_3,\vx_2),\\
%h_{11}(\vx_1,\vx_2)&=c_{11}(\vx_1,\vx_2)+
%\int\mathrm{d}\vx_3 c_{10}(\vx_1,\vx_3)\rho_0(\vx_3)h_{01}(\vx_3,\vx_2)
%+\int \mathrm{d}\vx_3 c_\mathrm{c}(\vx_1,\vx_3)\rho_1(\vx_3)h_{11}(\vx_3,\vx_2)
%+\int \mathrm{d}\vx_3 c_\mathrm{b}(\vx_1,\vx_3)\rho_1(\vx_3)h_\mathrm{c}(\vx_3,\vx_2).\\
\label{ec:OZ00}
h_{00}(\vx_1,\vx_2)=&\,c_{00}(\vx_1,\vx_2)+ (c_{00} \rho_0 \otimes h_{00})(\vx_1,\vx_2),\\
\label{ec:OZ10}
h_{10}(\vx_1,\vx_2)=&\,c_{10}(\vx_1,\vx_2)+ (c_{10} \rho_0 \otimes h_{00})(\vx_1,\vx_2)
+ (c_\mathrm{c} \rho_1 \otimes h_{10})(\vx_1,\vx_2),\\
\label{ec:OZ01}
h_{01}(\vx_1,\vx_2)=&\,c_{01}(\vx_1,\vx_2)+ (c_{00} \rho_0 \otimes h_{01})(\vx_1,\vx_2)
+ (c_{01} \rho_1 \otimes h_\mathrm{c})(\vx_1,\vx_2),\\
\label{ec:OZ11}
h_{11}(\vx_1,\vx_2)=&\,c_{11}(\vx_1,\vx_2)+ (c_{10} \rho_0 \otimes h_{01})(\vx_1,\vx_2)
+ (c_\mathrm{c} \rho_1 \otimes h_{11})(\vx_1,\vx_2) \nonumber\\
&+ (c_\mathrm{b} \rho_1 \otimes h_\mathrm{c})(\vx_1,\vx_2),\\
\label{ec:OZc}
h_\mathrm{c}(\vx_1,\vx_2)=&\,c_\mathrm{c}(\vx_1,\vx_2)+
(c_\mathrm{c} \rho_1 \otimes h_\mathrm{c})(\vx_1,\vx_2),
\end{align} 
where $(c\rho\otimes h)(\vx_1,\vx_2)\equiv \int \mathrm{d}\vx_3\,
c(\vx_1,\vx_3)\rho(\vx_3)h(\vx_3,\vx_2)$,
the subscripts $0$ and $1$ refer to the matrix and fluid, respectively, and
\begin{align}
\label{eq:h11}
h_{11}(\vx_1,\vx_2)&=h_\mathrm{c}(\vx_1,\vx_2)+h_\mathrm{b}(\vx_1,\vx_2),\\
\label{eq:c11}
c_{11}(\vx_1,\vx_2)&=c_\mathrm{c}(\vx_1,\vx_2)+c_\mathrm{b}(\vx_1,\vx_2),
\end{align}
where the subscripts c and b denote the \emph{connected} and \emph{blocking}
parts, respectively, of the correlation functions. In terms of the replicated
system, the blocking parts, $h_\mathrm{b}$ and $c_\mathrm{b}$, are the zero-replica
limit of the corresponding correlation functions between two different replicas
of the fluid \cite{given:1992a}. Clearly $h_{10}(\vx_1,\vx_2)=h_{01}(\vx_2,\vx_1)$
and $c_{10}(\vx_1,\vx_2)=c_{01}(\vx_2,\vx_1)$ and then it can be shown that
Eqs.~(\ref{ec:OZ10}) and (\ref{ec:OZ01}) are equivalent, so that 
Eqs.~(\ref{ec:OZ00}), (\ref{ec:OZ10}), (\ref{ec:OZ11}) and (\ref{ec:OZc}) form
an independent set.

Rosinberg et al.~\cite{rosinberg:1994} used the same replica trick to derive
the thermodynamics of these quenched-annealed (QA) systems. There are two important
results drawn from this work that concern the present paper: (i) the thermodynamics
is completely determined by the connected parts of the correlation functions,
and (ii) the connected and blocking parts of $h_{11}$ can be written without
any reference to replicas as
\begin{align}
\label{eq:hcnorepl}
\rho_1(\vx_1) \rho_1(\vx_2) h_\mathrm{c}(\vx_1,\vx_2)&=\rho_{11}(\vx_1,\vx_2)-
\overline{\rho(\vx_1|\{\vq_i\})\rho(\vx_2|\{\vq_i\})},\\
\label{eq:hbnorepl}
\rho_1(\vx_1) \rho_1(\vx_2) h_\mathrm{b}(\vx_1,\vx_2)&=
\overline{\rho(\vx_1|\{\vq_i\})\rho(\vx_2|\{\vq_i\})}-\rho_1(\vx_1)\rho_1(\vx_2),
\end{align}
where $\overline{\cdots}$ denotes the quenched average, $\rho(\vx|\{\vq_i\})$ is
the equilibrium density profile of the fluid for a particular configuration $\{\vq_i\}$ of
the disorder, $\rho_1(\vx)=\overline{\rho(\vx|\{\vq_i\})}$, and $\rho_{11}(\vx_1,\vx_2)$
is the disorder-averaged pair correlation function of the fluid.

The works of Madden and Glandt~\cite{madden:1988}, Given and Stell~\cite{given:1992a}
and Rosinberg et al.~\cite{rosinberg:1994} established the extension of the classical integral
equation theory to fluids in quenched disordered matrices. Since then, this has been
the main method to study QA systems and with its help much insight on
the phase behaviour of these systems have been gained. But the replica trick is closely
linked to the ROZ equations and so it has the typical limitations of any integral
equation theory: it is virtually impossible to apply the theory to non-uniform
phases. In the case of fluids without disorder this problem was solved by density
functional theories (DFTs), so it seems natural
to ask for an extension of DFT to QA systems.

There have been attempts to apply DFT to fluids in random media. For instance,
Menon and Dasgupta \cite{menon:1994} have constructed a Ramakrishnan-Yussouff density
functional, using the same replica trick employed in the
derivation of the ROZ equations, to study the effect of pinning in the freezing
of superconductor vortex lines.
The same approach has been applied to study hard spheres in a quenched random
gaussian potential \cite{thalmann:2000}. More recently, Schmidt~\cite{schmidt}
has proposed a DFT for QA mixtures also based on the replica trick.

In Schmidt's formalism, which we will refer to as replica-DFT (or simply rDFT),
the matrix is described by the equilibrium free-energy
density functional corresponding to the hamiltonian modelling the matrix particles,
while the behaviour of the fluid is ruled by the quenched-averaged grand potential
of the QA system $\Omega_\mathrm{rDFT}[\rho_1;\rho_0]$, which is written as a
functional of the disorder-average
density profile of the fluid, $\rho_1(\vx)$, and of the density profile
of the matrix, $\rho_0(\vq)$ (which enters as a parameter). The QA character
becomes explicit in the minimization principle imposed over the quenched-average
grand potential, which reads as
\begin{equation}
\frac{\delta\Omega_\mathrm{rDFT}[\rho_1;\rho_0]}{\delta \rho_1(\vx)}=0,
\end{equation}
where $\rho_0(\vq)$ is determined by the equation
\begin{equation}
\frac{\delta F_0[\rho_0]}{\delta \rho_0(\vq)}=u_0(\vq),
\end{equation}
$F_0[\rho_0]$ being the equilibrium free-energy functional of the pure matrix
and $u_0(\vq)\equiv \mu_0-\varphi_0(\vq)$, with $\mu_0$ the chemical
potential of the matrix and $\varphi_0(\vq)$ the external potential
acting over the matrix particles.

{}From $\Omega_\mathrm{rDFT}[\rho_1;\rho_0]$, the free-energy functional can be
defined as usual as
\begin{equation}
\Omega_\mathrm{rDFT}[\rho_1;\rho_0]=F^{\mathrm{id}}[\rho_1]+
F^\mathrm{ex}_\mathrm{rDFT}[\rho_1;\rho_0]-\int\mathrm{d}\vx u_1(\vx) \rho_1(\vx),
\end{equation}
where $F^{\mathrm{id}}[\rho_1]=kT\int\mathrm{d}\vx \rho_1(\vx)[\ln \mathcal{V}_1
\rho_1(\vx)-1]$ is the ideal contribution ($\mathcal{V}_1$ being the thermal volume
of the fluid particles) and $u_1(\vx)\equiv \mu_1-\varphi_1(\vx)$, $\mu_1$ being
the chemical potential of the fluid and $\varphi_1(\vx)$ the external potential
on the fluid particles. The excess contribution
$F^\mathrm{ex}_\mathrm{rDFT}[\rho_1;\rho_0]$
describes the interparticle interactions between fluid particles, and that
between fluid and matrix particles.

From a practical point of view, the functional
$F^\mathrm{ex}_\mathrm{rDFT}[\rho_1,\rho_0]$ should be approximated.
%It is important to note that the
%complete free-energy functional, $F_\mathrm{rDFT}[\rho_1;\rho_0]\equiv F^\mathrm{id}[\rho_1]
%+F^\mathrm{ex}_\mathrm{rDFT}[\rho_1,\rho_0]$, coincides with the quenched-averaged
%\begin{equation}
%F_\mathrm{rDFT}[\rho_1;\rho_0]=\overline{F[\rho_1|\{\vq_i\}]}+\overline{\int\mathrm{d}\vx
%\, \sum_{i=1}^M \varphi_{10}(\vx,\vq_i) \rho_1(\vx|\{\vq_i\})},
%\end{equation}
%where $F[\rho_1|\{\vq_i\}]$ is the intrinsic free-energy functional of the pure
%fluid system under the action of the external potential
%$\varphi^\mathrm{ext}_1(\vx)+\sum_{i=1}^M \varphi_{10}(\vx,\vq_i)$,
%where the second term represents the contribution due to the interaction with the
%matrix particles fixed at positions $\{\vq_1,\dotsc,\vq_M\}$.
Schmidt's proposal for $F^\mathrm{ex}_\mathrm{rDFT}$ is based on fundamental measure
theory~\cite{rosenfeld:1989,tarazona:1997,tarazona:2000}.
This approximation has been applied
to study the phase behaviour of colloid-polymer mixtures in bulk random matrices,
of rods in quenched sphere matrices, of spheres in random fibre networks, and of
soft-core fluids in soft-core matrices~\cite{schmidt:rDFTapp}.
Also, with the lattice version of fundamental measure theory~\cite{lafuente:LFMT},
it has been applied
to study the freezing transition in a hard-core discrete fluid with different
kinds of matrices~\cite{schmidt:2003}. Thus, as shown by its
applications, rDFT is an important step forward in the study of QA systems.

This notwithstanding, the theory has a number of weak points which should be
pointed out. Although the replica trick is a widely-applied method of statistical
physics, it makes a few assumptions which are difficult to
justify concerning the analytic continuation of the
grand potential as a function of the number of replicas, and
the replica symmetry or its breaking. Hence an alternative derivation
of DFT for QA systems would be desirable. Moreover, contrary to what happens in
classical DFT, the formulation of rDFT makes it difficult to derive
the set of OZ equations for QA systems from functional relations. As a matter
of fact, at present it is not at all clear what the meaning of the second
derivatives of $F^\mathrm{ex}_\mathrm{rDFT}[\rho_1;\rho_0]$ is. These problems
are the two main motivations of this paper.

The remaining of the paper is organized as follows. In section~2, we propose a DFT
for QA systems based on the convexity properties
of the quenched-averaged grand potential. The derivation of the formalism resembles
that of classical DFT and makes no use of the replica trick.
We will Legendre-transform the grand potential to obtain the quenched-average
``intrinsic'' free-energy functional, $\mathcal{F}[\rho_1,\rho_{10};\mathcal{Q}]$,
which depends on the quenched-averaged density profile of the fluid, $\rho_1(\vx)$,
on the pair distribution function of the fluid and matrix particles,
$\rho_{10}(\vx,\vq)$, and on the probability distribution of
disorder $\mathcal{Q}$. The dependence on $\rho_{10}(\vx,\vq)$ can be eliminated to
obtain a functional only of $\rho_1(\vx)$ (for fixed fluid-matrix interaction),
which will play the same role as the standard free-energy functional in
classical DFT, and coincides with Schmidt's $F_\mathrm{rDFT}[\rho_1;\rho_0]$.
We will then proceed with
one of the most important contributions of this work: the derivation
of a set of OZ equations where the direct correlation functionals are identified with
second derivatives of this functional. Only equations~(\ref{ec:OZ00}--\ref{ec:OZ01})
and (\ref{ec:OZc}) can be derived within this DFT approach, but as we will discuss,
these form a closed set of equations which involve all the structure information
that is relevant to the thermodynamics of the system. Section~2 concludes showing
how to derive the thermodynamics within this DFT approach. In section~3,
we will make this formalism explicit in the case of an ideal fluid adsorbed in an
arbitrary matrix. Conclusions and further discussions are gathered
in section~4.

\section{Quenched-annealed density functional theory}

The system that we aim to describe consists of a fluid inside a porous
matrix with which it interacts. The matrix is formed by a distribution of
particles quenched at positions ${\vq}_i$, $i=1,\dots,M$. The fluid 
consists of particles, whose positions are denoted ${\vx}_i$, $i=1,\dots,N$,
and whose interactions are described by the hamiltonian (divided by $kT$)
${\cal H}_N({\vx}_1,\dots,{\vx}_N)$. These particles are in equilibrium
with a thermal bath at chemical potential $\mu_1$ and
each of them undergoes the action of an external potential $\varphi_1({\vx})$.
Besides, a fluid particle at position ${\vx}$ interacts with a matrix
particle at position ${\vq}$ through the interaction potential
$\varphi_{10}({\vx},{\vq})$. To all purposes, the total external potential
acting on a fluid particle at position ${\vx}$ is
\[
V_{\rm ext}({\vx})=\varphi_1({\vx})+\sum_{i=1}^M\varphi_{10}({\vx},{\vq}_i).
\]
The grand partition function for this system will be ($\beta=1/kT$)
\begin{equation}
\begin{split}
\Xi\big[u_1,u_{10}|\{{\vq}_i\}\big]=&\, 1+\sum_{N=1}^{\infty}\frac{1}{{\cal V}_1^NN!}
\int \mathrm{d}{\vx}_1\cdots \mathrm{d}{\vx}_N\,
\exp\bigg\{-{\cal H}_N({\vx}_1,\dots,{\vx}_N) \\
&+\beta\sum_{i=1}^N\bigg[u_1({\vx}_i)+\sum_{j=1}^Mu_{10}({\vx}_i,{\vq}_j)
\bigg]\bigg\},
\end{split}
\end{equation}
which is a functional of $u_1({\vx})\equiv\mu_1-\varphi_1({\vx})$ and
$u_{10}({\vx},{\vq})\equiv -\varphi_{10}({\vx},{\bf q})$, and also depends
on the set $\{{\vq}_i\}$. Accordingly, the grand potential will be
\begin{equation}
\Omega\big[u_1,u_{10}|\{{\vq}_i\}\big]=-kT\ln\Xi\big[u_1,u_{10}|\{{\bf q}_i\}\big].
\end{equation}

Now we need a model for the porous matrix. The simplest model is to assume
that matrix particles are placed at random positions, according to a 
probability density \cite{madden:1988}.
Thus the grand potential is a random variable. The
hypothesis we make now is that the grand potential per unit volume, in
the thermodynamic limit, is a self-averaging random variable; therefore
we can obtain its value in this limit by simply averaging over disorder (matrix
particle positions). Hence the grand potential for the system in the thermodynamic
limit is obtained as
\begin{equation}
\Omega[u_1,u_{10};\mathcal{Q}]\equiv\overline{\Omega\big[u_1,u_{10}|\{{\vq}_i\}\big]},
\label{eq:grandpotential}
\end{equation}
where $\mathcal{Q}(\{\vq_i\})$ is the probability density of the matrix positions,
and $\overline{\,\cdots\,}$ denotes a $\mathcal{Q}$-average. This puts the
quenched average into play.

\subsection{Concavity of the grand potential}

It is convenient to introduce the state functions
\begin{equation}
\hat\rho_N({\vx})\equiv\sum_{i=1}^N\delta({\vx}-{\vx}_i),\qquad
\hat\rho^{0}_M({\vq})\equiv\sum_{i=1}^M\delta({\bf q}-{\bf q}_i).
\end{equation}
In terms of them
\begin{eqnarray}
\Xi\big[u_1,u_{10}|\{{\vq}_i\}\big] &=& 1+\sum_{N=1}^{\infty}\frac{1}{{\cal V}_1^NN!}
\int\mathrm{d}{\vx}_1\cdots \mathrm{d}{\vx}_N\,
\exp\bigg\{-{\cal H}_N({\vx}_1,\dots,{\vx}_N) \nonumber \\
&&+\beta\left\langle u_1,\hat\rho_N\right\rangle+\beta\left\langle u_{10},
\hat\rho_N\hat\rho^{0}_M\right\rangle\bigg\},
\end{eqnarray}
where
\begin{equation}
\left\langle u_1,\hat\rho_N\right\rangle\equiv\int\mathrm{d}{\vx}\,
u_1({\vx})\hat\rho_N({\vx}),\qquad
\left\langle u_{10},\hat\rho_N\hat\rho^{0}_M\right\rangle\equiv
\int\mathrm{d}{\vx}\mathrm{d}{\vq}\,
u_{10}({\vx},{\vq})\hat\rho_N({\vx})\hat\rho^{0}_M({\bf q}).
\end{equation}
With these definitions it is straightforward that
\begin{equation}
\begin{split}
-\frac{\delta\Omega[u_1,u_{10};\mathcal{Q}]}{\delta u_1({\vx})} &=\overline{\rho\big({\vx}|
\{{\vq}_i\}\big)}=\rho_1({\vx}), \\
-\frac{\delta\Omega[u_1,u_{10};\mathcal{Q}]}{\delta u_{10}({\vx},{\vq})} &
=\overline{\rho\big({\vx}|\{{\vq}_i\}\big)\hat\rho^{0}_M({\bf q})}=
\rho_{10}({\vx},{\vq}),
\end{split}
\label{eq:densities}
\end{equation}
where $\rho\big({\vx}|\{{\vq}_i\}\big)$ denotes the equilibrium
density of the fluid for fixed positions of the matrix particles, and
$\rho_1({\vx})$ is the quenched-averaged density profile
of the fluid. Likewise, $\rho_{10}({\vx},{\vq})$ is the pair correlation
function of the fluid and matrix particles.

On the other hand, $\Omega[u_1,u_{10};\mathcal{Q}]$ is a concave functional
of both, $u_1$ and $u_{10}$. This is easily proven by evaluating
$\Xi\big[u_1,u_{10}|\{{\vq}_i\}\big]$ on
$u_1^{(\lambda)}({\vx})=\lambda u_1^{(1)}({\vx})
+(1-\lambda)u_1^{(0)}({\vx})$ ($0<\lambda<1$).
Since
\[
\exp\{\langle u_1^{(\lambda)},\hat\rho_N\rangle\}=\left(
\exp\{\langle u_1^{(1)},\hat\rho_N\rangle\}\right)^{\lambda}
\left(\exp\{\langle u_1^{(0)},\hat\rho_N\rangle\}\right)^{1-\lambda},
\]
by H\"older's inequality
\cite{caillol:2002} we get
\[
\Xi\left[u_1^{(\lambda)},u_{10}\Big|\{{\vq}_i\}\right]<
\Xi\left[u_1^{(1)},u_{10}\Big|\{{\vq}_i\}\right]^{\lambda}
\Xi\left[u_1^{(0)},u_{10}\Big|\{{\vq}_i\}\right]^{1-\lambda},
\]
from which
\[
\Omega\left[u_1^{(\lambda)},u_{10}\Big|\{{\vq}_i\}\right]>
\lambda\Omega\left[u_1^{(1)},u_{10}\Big|\{{\vq}_i\}\right]+
(1-\lambda)\Omega\left[u_1^{(0)},u_{10}\Big|\{{\vq}_i\}\right],
\]
and averaging over disorder,
\[
\Omega\left[u_1^{(\lambda)},u_{10};\mathcal{Q}\right]>
\lambda\Omega\left[u_1^{(1)},u_{10};\mathcal{Q}\right]+
(1-\lambda)\Omega\left[u_1^{(0)},u_{10};\mathcal{Q}\right].
\]
Clearly the same holds for $u_{10}$. Because of this, equations
(\ref{eq:densities}) define a one-to-one correspondence between
the pair $\{\rho_1,\rho_{10}\}$ and the pair $\{u_1,u_{10}\}$
(i.e.\ the equations can be inverted)
\cite{caillol:2002}.

\subsection{Free-energy functional and minimum principle}

Let us now introduce the Legendre transform of $\Omega[u_1,u_{10};\mathcal{Q}]$
with respect to its two arguments
\begin{equation}
\mathcal{F}\left[\rho_1,\rho_{10};\mathcal{Q}\right]\equiv\Omega[u_1,u_{10};\mathcal{Q}]
+\langle u_1,\rho_1\rangle +\left\langle u_{10},\rho_{10}\right\rangle,
\label{eq:legendre}
\end{equation}
where $u_1({\vx})$ and $u_{10}({\vx},{\vq})$ are the solution of
Eqs.~(\ref{eq:densities}) for fixed $\rho_1({\vx})$ and $\rho_{10}({\vx},{\vq})$.
Because of the properties of the Legendre transform \cite{caillol:2002}
\begin{enumerate}[(a)]
\item $\mathcal{F}\left[\rho_1,\rho_{10};\mathcal{Q}\right]$ is a {\em convex}
functional of both $\rho_1({\vx})$ and $\rho_{10}({\vx},{\vq})$;
\item the equilibrium $\rho_1({\vx})$ and $\rho_{10}({\vx},{\vq})$
are the absolute minimum of the functional
\begin{equation}
\widetilde\Omega\left[\rho_1,\rho_{10};\mathcal{Q}\right]\equiv
\mathcal{F}\left[\rho_1,\rho_{10};\mathcal{Q}\right]-\langle u_1,\rho_1\rangle-
\left\langle u_{10},\rho_{10}\right\rangle,
\end{equation}
for fixed $u_1({\vx})$ and $u_{10}({\vx},{\vq})$, and therefore
\item they can be obtained by solving the equations
\begin{equation}
\frac{\delta\mathcal{F}\left[\rho_1,\rho_{10};\mathcal{Q}\right]}{\delta\rho_1(\vx)}
 = u_1(\vx), \qquad
\frac{\delta\mathcal{F}\left[\rho_1,\rho_{10};\mathcal{Q}\right]}{\delta\rho_{10}(\vx,
\vq)}= u_{10}(\vx,\vq).
\label{eq:EL}
\end{equation}
\end{enumerate}

As for the meaning of the functional $\mathcal{F}$, let us adopt a different
point of view on the system: let us think of the porous matrix as an
external potential acting on the fluid particles. Then the ``intrinsic'' free-energy
functional is obtained as
\[\begin{split}
\mathcal{F}[\rho] &=\Omega\big[u_1,u_{10}|\{{\vq}_i\}\big]+
\int\mathrm{d}{\vx}\,\rho\big({\vx}|\{{\vq}_i\}\big)\left\{u_1({\vx})
+\sum_{i=1}^mu_{10}({\vx},{\vq}_i)\right\} \\
&=\Omega\big[u_1,u_{10}|\{{\vq}_i\}\big]+
\int\mathrm{d}{\vx}\,\rho\big({\vx}|\{{\vq}_i\}\big)u_1({\vx})+
\int\mathrm{d}{\vx}\mathrm{d}{\vq}\,\rho\big({\vx}|\{{\bf q}_i\}\big)\hat\rho_M^0({\bf q})
u_{10}({\vx},{\vq}),
\end{split}\]
and is, of course, a functional of $\rho\big({\vx}|\{{\vq}_i\}\big)$.
If we now average this functional over disorder we obtain, making use of
Eqs.~(\ref{eq:grandpotential}), (\ref{eq:densities}) and (\ref{eq:legendre}),
\begin{equation}
\overline{\mathcal{F}[\rho]}=\mathcal{F}[\rho_1,\rho_{10};\mathcal{Q}].
\end{equation}
This equation reveals the physical meaning of functional
$\mathcal{F}[\rho_1,\rho_{10};\mathcal{Q}]$ as the intrinsic
free energy of the fluid
undergoing the presence of a porous matrix, averaged over disorder. But
the fact that this functional depends on both $\rho_1(\vx)$ and
$\rho_{10}(\vx,\vq)$ makes it rather inconvenient to use it as the
basis for a QA-DFT (notice that this functional is the same for any
fluid-matrix interaction, so it is far too general).

Of course, one can assume $u_{10}(\vx,\vq)$ fixed and Legendre-transform
only with respect to $u_1(\vx)$ to obtain the alternative functional
\begin{equation}
\label{eq:psi}
F[\rho_1;\mathcal{Q}]=\Omega[u_1,u_{10};\mathcal{Q}]+\langle u_1,\rho_1\rangle.
\end{equation}
This is [for fixed $u_{10}(\vx,\vq)$ and $\mathcal{Q}$] a functional of
$\rho_1(\vx)$ alone, and fulfils the Euler-Lagrange equation
\begin{equation}
\label{eq:dual}
\frac{\delta F[\rho_1;\mathcal{Q}]}{\delta\rho_1(\vx)}=u_1(\vx).
\end{equation}
Comparing with (\ref{eq:legendre}) and recalling that $u_{10}(\vx,\vq)=
-\varphi_{10}(\vx,\vq)$,
\begin{equation}
 F[\rho_1;\mathcal{Q}]=\mathcal{F}[\rho_1,\rho_{10};\mathcal{Q}]+
\langle\varphi_{10},\rho_{10}\rangle,
\end{equation}
the intrinsic free energy of the fluid \emph{plus} the interaction energy
with the porous matrix. As we will show in Sec.~\ref{sec:ROZ}, this one and not 
$\mathcal{F}[\rho_1,\rho_{10};\mathcal{Q}]$ is the functional that
plays a similar role in QA-DFT as the standard free-energy functional does
in classical DFT, and in fact coincides with the functional
$F_\mathrm{rDFT}[\rho_1]$ derived from the replica formalism.

\subsection{Replica Ornstein-Zernike equations}
\label{sec:ROZ}

Let us work out the identity
\begin{equation}
\delta(\vx-\vx')=\frac{\delta\rho_1(\vx)}{\delta\rho_1(\vx')}=
\int\mathrm{d}\vy\,\frac{\delta\rho_1(\vx)}{\delta u_1(\vy)}\frac{\delta u_1(\vy)}{\delta
\rho_1(\vx')}=-\int\mathrm{d}\vy\,\frac{\delta^2\Omega[u_1,u_{10};\mathcal{Q}]}{\delta u_1(\vx)
\delta u_1(\vy)}\frac{\delta^2 F[\rho_1;\mathcal{Q}]}{\delta\rho_1(\vy)\rho_1(\vx')}.
\label{eq:delta}
\end{equation}
This is one of the replica Ornstein-Zernike (ROZ) equations, namely Eq.~(\ref{ec:OZc}).
To see it let us compute
\begin{eqnarray}
-kT\frac{\delta^2 \Omega[u_1,u_{10};\mathcal{Q}]}{\delta u_1(\vx) \delta u_1(\vy)} &=&
\overline{\rho_{11}(\vx,\vy|\{\vq_i\})-\rho_1(\vx|\{\vq_i\})\rho_1(\vy|\{\vq_i\})
-\delta(\vx-\vy)\rho_1(\vx|\{\vq_i\})} \nonumber\\
&=& \rho_1(\vx)\rho_1(\vy) h_\mathrm{c}(\vx,\vy)- \delta(\vx,\vy)\rho_1(\vx),
\label{eq:hc}
\end{eqnarray}
where we have made use of Eq.~(\ref{eq:hcnorepl}). Introducing this expression into
Eq.~(\ref{eq:delta})
we are immediately led to the identification
\begin{equation}
\label{eq:cc}
-\beta\frac{\delta^2 F^{\rm ex}[\rho_1;\mathcal{Q}]}{\delta\rho_1(\vx)\rho_1(\vx')}
=c_c(\vx,\vx';[\rho_1;\mathcal{Q}]),
\end{equation}
where $F^{\rm ex}[\rho_1;\mathcal{Q}]$ is the excess (over the ideal) part of
the functional $F[\rho_1;\mathcal{Q}]$.

In order to obtain Eqs.~(\ref{ec:OZ00}--\ref{ec:OZ01}) we must set $\mathcal{Q}$
as the probability distribution of 
a grand-canonical ensemble at temperature $T_0$, chemical
potential $\mu_0$ and external potential $\varphi_0(\vq)$.
Thus, if we define as usual $u_0(\vq)\equiv\mu_0-\varphi_0(\vq)$,
the probability of finding the matrix configuration $\{M;\vq_1,\dotsc,\vq_M\}$
is given by $\mathcal{Q}=\{P_M(\vq_1,\dotsc,\vq_M)\}_{M\ge 0}$, where
\begin{equation}
P_M(\vq_1,\dotsc,\vq_M)=\frac{1}{\Xi_0[u_0]}\frac{1}{\mathcal{V}_0^M M!}
\exp\left\{-\mathcal{H}^0_M(\vq_1,\dotsc,\vq_M)+\langle u_0,\hat\rho^0_M\rangle\right\},
\end{equation}
$\mathcal{H}^0_M(\vq_1,\dotsc,\vq_M)$ being the hamiltonian which models
the interaction between matrix particles (divided by $kT_0$) and $\Xi_0[u_0]$
the grand partition function
\begin{equation}
\Xi_0[u_0]=1+\sum_{M=1}^\infty \frac{1}{\mathcal{V}_0^M M!}\int\mathrm{d}\vq_1\dotsb
\mathrm{d}\vq_M \exp\left\{-\mathcal{H}^0_M(\vq_1,\dotsc,\vq_M)+\langle u_0,
\hat\rho^0_M\rangle\right\}
\end{equation}
($\mathcal{V}_0$ is the thermal volume of the matrix fluid). From classical DFT
we know that for each external potential $u_0(\vq)$ there
exists a unique equilibrium density profile $\rho_0(\vq)$ and the system
can be described alternatively in terms of any of them. Therefore, if the hamiltonian
$\mathcal{H}^0_M(\vq_1,\dotsc,\vq_M)$ and temperature $T_0$ remain fixed, the
dependence of the functionals $\Omega[u_1;\mathcal{Q}]$ and
$F[\rho_1;\mathcal{Q}]$ on the disorder
is actually a dependence on either $u_0(\vq)$ or $\rho_0(\vq)$. Hereafter, we will
make explicit this dependence by writing this functional as $F[\rho_1;\rho_0]$
and the grand potential as $\Omega[u_1;u_0]$.

Now, Eq.~(\ref{ec:OZ00}) is just the identity
\begin{equation}
\delta(\vq-\vq')=\frac{\delta \rho_0(\vq)}{\delta \rho_0(\vq')}=
\int \mathrm{d}\vs\,\frac{\delta\rho_0(\vq)}{\delta u_0(\vs)}\frac{\delta u_0(\vs)}{\delta
\rho_0(\vq')}=-\int\mathrm{d}\vs\,\frac{\delta^2\Omega_0[u_0]}{\delta u_0(\vq)
\delta u_0(\vs)}\frac{\delta^2 F_0[\rho_0]}{\delta\rho_0(\vs)\rho_0(\vq')},
\end{equation}
where $\Omega_0[u_0]=-kT_0\ln\Xi_0[\rho_0]$ is the grand potential of the
matrix and $F_0[\rho_0]$ the corresponding free-energy functional.
Finally, to obtain Eqs.~(\ref{ec:OZ10}) and (\ref{ec:OZ01}), we will notice
that the QA system can be described in terms of any of the following
pairs of independent functions $\{u_1(\vx),u_0(\vq)\}$ or $\{\rho_1(\vx),\rho_0(\vq)\}$.
Then, both ROZ equations can be identified, respectively, with the identities
\begin{align}
0=\frac{\delta u_1(\vx)}{\delta u_0(\vq)}&=\int\mathrm{d}\vy\,
\frac{\delta u_1(\vx)}{\delta \rho_1(\vy)}\frac{\delta \rho_1(\vy)}{\delta u_0(\vq)}+
\int\mathrm{d}\vs\, \frac{\delta u_1(\vx)}{\delta\rho_0(\vs)}
\frac{\delta \rho_0(\vs)}{\delta u_0(\vq)} \nonumber \\
&=-\int\mathrm{d}\vy\,\frac{\delta^2 F[\rho_1;\rho_0]}
{\delta \rho_1(\vx) \delta \rho_1(\vy)}\frac{\delta^2 \Omega[u_1;u_0]}{\delta u_1(\vy)
\delta u_0(\vq)}-
\int\mathrm{d}\vs\, \frac{\delta^2 F[\rho_1;\rho_0]}{\delta \rho_1(\vx)\delta\rho_0(\vs)}
\frac{\delta^2 \Omega_0[u_0]}{\delta u_0(\vs) \delta u_0(\vq)},\\[3mm]
0=\frac{\delta \rho_1(\vx)}{\delta \rho_0(\vq)}&=-\int\mathrm{d}\vy\,\frac{\delta
\rho_1(\vx)}{\delta u_1(\vy)}\frac{\delta u_1(\vy)}{\delta \rho_0(\vq)}-
\int\mathrm{d}\vs\, \frac{\delta\rho_1(\vx)}{\delta u_0(\vs)}
\frac{\delta u_0(\vs)}{\delta \rho_0(\vq)} \nonumber \\
&=-\int\mathrm{d}\vy\,\frac{\delta^2 \Omega[u_1;u_0]}
{\delta u_1(\vx) \delta u_1(\vy)}\frac{\delta^2 F[\rho_1;\rho_0]}{\delta \rho_1(\vy)
\delta \rho_0(\vq)}-
\int\mathrm{d}\vs\, \frac{\delta^2 \Omega[u_1;u_0]}{\delta u_1(\vx)\delta u_0(\vs)}
\frac{\delta^2 F_0[u_0]}{\delta \rho_0(\vs) \delta \rho_0(\vq)},
\end{align}
where we have used Eqs.~(\ref{eq:densities}) and (\ref{eq:dual}) for the
QA system, and their counterparts for the matrix. To complete the identification
of these identities with the corresponding OZ equations, we have to take into account
the expression (\ref{eq:hc}) as well as
\begin{equation}
-kT\frac{\delta^2 \Omega[u_1;u_0]}{\delta u_1(\vx) \delta u_0(\vq)}=
\rho_1(\vx)\rho_0(\vq)h_{10}(\vx,\vq),
\end{equation}
and to make the identification
\begin{equation}
\label{eq:c10}
-\beta\frac{\delta^2 F[\rho_1;\rho_0]}{\delta \rho_1(\vx)\delta\rho_0(\vq)}=
c_{10}(\vx,\vq;[\rho_1;\rho_0]).
\end{equation}

At this point, it is important to notice that the set of OZ equations that
we have obtained within the DFT approach is self-contained. This means
that if we have the functionals $F_0[\rho_0]$ and $F[\rho_1;\rho_0]$, we can derive
from them the direct correlation functionals $c_{00}$, $c_\mathrm{c}$
and $c_{10}$, and using them as inputs in the OZ
Eqs.~(\ref{ec:OZ00}), (\ref{ec:OZ10}) and (\ref{ec:OZc}) to obtain
$h_{00}$, $h_{10}$ and $h_\mathrm{c}$.
Moreover, both Eqs.~(\ref{ec:OZ00}) and (\ref{ec:OZc}) can be solved
independently and their solutions can be used to solve Eq.~(\ref{ec:OZ10}).
This situation is remarkably different from the one we find in the integral
equation framework. In that case, although the OZ equation for
the matrix [Eq.~(\ref{ec:OZ00})] is independent of all the others,
the remaining ones [Eqs.~(\ref{ec:OZ10}), (\ref{ec:OZ11}) and (\ref{ec:OZc})]
form a coupled system.
The reason for this difference is that in the case of integral equation
theory, the direct correlation functions are also unknown and the ROZ
equations must be complemented with closure relations.
This additional equations are derived from exact
relations between the interaction potentials between particles
and the correlation functions, with one equation for each
potential. Thus, in our case, we would have two new equations which would involve
$\{h_{10},c_{10}\}$ and $\{h_{11},c_{11}\}$, respectively. The absence
of a closure relation for $\{h_\mathrm{c},c_\mathrm{c}\}$ is what
keeps the set of Eqs.~(\ref{ec:OZ10}), (\ref{ec:OZ11}) and (\ref{ec:OZc}),
the two closure relations, and one of Eqs.~(\ref{eq:h11}) or (\ref{eq:c11}),
coupled.

We should remark that in the QA-DFT the blocking parts
are absent. Nevertheless, contrary to what happens with integral equation theory,
we are able to compute all the structure functions that are relevant to the
thermodynamics without the blocking correlations.

\subsection{Thermodynamics}

We will finish this section showing how all the thermodynamics can
be derived from the functional $F[\rho_1;\rho_0]$. The starting point will
be the relation proved by Rosinberg et al.~\cite{rosinberg:1994}
\begin{equation}
\overline{\Omega[u_1|\{\vq_i\}]}=\Omega[u_1;u_0]=-pV,
\end{equation}
where $p$ is the thermodynamic pressure and $V$ the volume of the system
(see Refs.~\cite{dong:2005} and \cite{kierlik:1995} for a
discussion about the definition of the thermodynamic pressure
and the difference between this one and the mechanical pressure).
Now, as the functional $F[\rho_1;\rho_0]$ is related to the quenched-averaged grand
potential $\Omega[u_1;u_0]$ [Eq.~(\ref{eq:psi})] in the same way as the standard
free-energy functional with the grand potential in classical DFT,
and as this formal equivalence is also found in the relation (\ref{eq:dual})
between $F[\rho_1;\rho_0]$ and the chemical potential, we can conclude that
all the thermodynamic relations we found in classical DFT
remain formally identical in the QA-DFT, with
$c_\mathrm{c}$ playing the role of the direct correlation because of
Eq.~(\ref{eq:cc}).

\section{An exact model: ideal fluid in an arbitrary matrix}

As the simplest example let us consider the only known example which
can be exactly solved in this formalism: an ideal fluid in an arbitrary
porous matrix. As a QA system, the matrix is taken to be a configuration of
a grand-canonical ensemble of a certain fluid at temperature $T_0$, chemical
potential $\mu_0$, and external potential $\varphi_0({\vq})$ [let us also define
$u_0({\vq})\equiv\mu_0-\varphi_0({\bf q})$]. The grand partition function, grand
potential and free energy of this fluid will be denoted, respectively,
$\Xi_0[u_0]$, $\Omega_0[u_0]$ and $F_0[\rho_0]$, $\rho_0({\vq})$ being the
corresponding equilibrium density profile.

For the ideal gas $\mathcal{H}_N=0$, so the grand partition function of the
fluid becomes
\begin{equation}
\begin{split}
\Xi\big[u_1,u_{10}|\{{\vq}_i\}\big] &=1+\sum_{N=1}^{\infty}\frac{1}{\mathcal{V}_1^N
N!}\left\{\int\mathrm{d}{\vx}\,\exp\left[\beta u_1({\vx})+\sum\limits_{i=1}^M\beta
u_{10}({\vx},{\vq}_i)\right]\right\}^N \\
&=\exp\left\{\frac{1}{\mathcal{V}_1}\int\mathrm{d}{\vx}\,\exp\left[\beta u_1({\vx})
+\sum\limits_{i=1}^M\beta u_{10}({\vx},{\vq}_i)\right]\right\}.
\end{split}
\end{equation}
Thus,
\begin{equation}
\Omega\big[u_1,u_{10}|\{{\vq}_i\}\big]=-\frac{kT}{\mathcal{V}_1}\int\mathrm{d}{\vx}\,
\exp\left[\beta u_1({\vx})+\sum\limits_{j=1}^M\beta u_{10}({\vx},{\vq}_i)\right],
\end{equation}
and therefore
\begin{eqnarray}
\Omega[u_1,u_{10};\mathcal{Q}] &=& -\frac{kT}{\mathcal{V}_1}\int\mathrm{d}{\vx}\,
e^{\beta u_1({\vx})}\frac{\Xi_0[\tilde u_0({\vx},\cdot)]}{\Xi_0[u_0]} \nonumber\\
&=& -\frac{kT}{\mathcal{V}_1}\int\mathrm{d}{\vx}\,
e^{\beta u_1({\vx})-\beta_0\Delta\Omega_0({\vx})}.
\label{eq:Omega}
\end{eqnarray}
In this expressions $\Xi_0\big[\tilde u_0(\vx,\cdot)\big]$ stands for the
grand partition function of the matrix fluid undergoing an external potential 
$\tilde u_0(\vx,\vq)\equiv u_0({\vq})+(T_0/T)u_{10}({\vx},{\bf q})$,
$\vx$ being the position of
a fixed fluid particle, and $\Delta\Omega_0({\vx})\equiv\Omega_0\big[\tilde
u_0({\vx},\cdot)\big]-\Omega_0[u_0]$.

{}From the first of Eqs.~(\ref{eq:densities}) it follows that
\begin{equation}
\rho_1(\vx)=\frac{1}{\mathcal{V}_1}e^{\beta u_1({\vx})-\beta_0\Delta\Omega_0({\vx})},
\label{eq:rho1}
\end{equation}
an interesting equation which tells us that the average equilibrium density
profile of the fluid is given by the barometric law corrected with the
probability of inserting a fluid particle in the matrix fluid at position
$\vx$, namely $e^{-\beta_0\Delta\Omega_0({\vx})}$.

{}From the second of Eqs.~(\ref{eq:densities}) it follows that
\begin{equation}
\rho_{10}(\vx,\vq)=\frac{1}{\mathcal{V}_1}e^{\beta u_1({\vx})}
\frac{kT_0}{\Xi_0[u_0]}\frac{\delta\Xi_0[\tilde u_0(\vx,\cdot)]}
{\delta \tilde u_0(\vx,\vq)}.
%\label{eq:rho10}
\end{equation}
Dividing this equation by Eq.~(\ref{eq:rho1}) leads to
\begin{equation}
\frac{\rho_{10}(\vx,\vq)}{\rho_1(\vx)}=-\frac{\delta\Omega_0[\tilde u_0(\vx,\cdot)]}
{\delta \tilde u_0(\vx,\vq)}=\rho_0(\vx,\vq),
\label{eq:rho10bis}
\end{equation}
where $\rho_0(\vx,\vq)$ is the equilibrium density profile of the
matrix fluid corresponding to the external potential $\tilde u_0(\vx,\vq)$
created by a fluid particle placed at $\vx$.

Because of Eq.~(\ref{eq:rho1}), Eq.~(\ref{eq:Omega}) simply becomes
\begin{equation}
\Omega[u_1,u_{10};\mathcal{Q}]=-kT\int\mathrm{d}\vx\,\rho_1(\vx),
\end{equation}
the equation of state of the ideal gas. On the other hand, eliminating $u_1(\vx)$
from Eq.~(\ref{eq:rho1}),
\begin{equation}
u_1(\vx)=kT\ln\Big(\mathcal{V}_1\rho_1(\vx)\Big)+\frac{T}{T_0}\Delta\Omega_0(\vx),
\end{equation}
so Eq.~(\ref{eq:legendre}) becomes
\begin{equation}
\begin{split}
\mathcal{F}[\rho_1,\rho_{10};\mathcal{Q}]=&\,F^{\rm id}[\rho_1]+\frac{T}{T_0}
\int\mathrm{d}\vx\,\Big\{\rho_1(\vx)\Omega_0[\tilde u_0(\vx,\cdot)]-
\rho_1(\vx)\Omega_0[u_0]\Big\} \\
&+\int\mathrm{d}\vx\int\mathrm{d}\vq\,u_{10}(\vx,\vq)\rho_{10}(\vx,\vq),
\end{split}
\label{eq:FOmega}
\end{equation}
where $u_{10}(\vx,\vq)$ is the solution to Eq.~(\ref{eq:rho10bis}) and
\begin{equation}
F^{\rm id}[\rho_1]=kT\int\mathrm{d}\vx\,\rho_1(\vx)
\left\{\ln\Big(\mathcal{V}_1\rho_1(\vx)\Big)-1\right\}.
\end{equation}
Adding and subtracting
\[
\frac{T}{T_0}\int\mathrm{d}\vx\,\rho_1(\vx)\left\{
\int\mathrm{d}\vq\,u_0(\vq)\Big(\rho_0(\vq)-\rho_0(\vx,\vq)\Big)\right\}
\]
to Eq.~(\ref{eq:FOmega}) and using Eq.~(\ref{eq:rho10bis}) yields
\begin{equation}
\begin{split}
\mathcal{F}[\rho_1,\rho_{10};\mathcal{Q}]
=&\,F^{\rm id}[\rho_1]+\frac{T}{T_0}\int\mathrm{d}\vx\,\rho_1(\vx)\left\{
\int\mathrm{d}\vq\,u_0(\vq)\Big(\rho_0(\vq)-\rho_0(\vx,\vq)\Big)\right\} \\
&+\frac{T}{T_0}\int\mathrm{d}\vx\,\rho_1(\vx)\left\{
\Omega_0[\tilde u_0({\vx},\cdot)]
+\int\mathrm{d}\vq\,\tilde u_0(\vx,\vq)\rho_0(\vx,\vq) \right\} \\
&-\frac{T}{T_0}\int\mathrm{d}\vx\,\rho_1(\vx)\left\{
\Omega_0[u_0]+\int\mathrm{d}\vq\,u_0(\vq)\rho_0(\vx)\right\}.
\end{split}
\end{equation}
One can recognize in the brackets above the Legendre transforms
of the grand potential of the matrix fluid; thus the final expression for
the functional can be written as
\begin{equation}
\begin{split}
\mathcal{F}[\rho_1,\rho_{10};\mathcal{Q}] =&\,
F^{\rm id}[\rho_1]
+\frac{T}{T_0}\int\mathrm{d}\vx\,\rho_1(\vx)\Big\{
F_0\left[\rho_{10}(\vx,\cdot)/\rho_1(\vx)\right]-F_0[\rho_0]\Big\} \\
&+\frac{T}{T_0}\int\mathrm{d}\vx\int\mathrm{d}\vq\,
\frac{\delta F_0[\rho_0]}{\delta\rho_0(\vq)}
\Big\{\rho_1(\vx)\rho_0(\vq)-\rho_{10}(\vx,\vq)\Big\}.
\end{split}
\end{equation}
%where we have set
%\begin{equation}
%u_0(\vq)=\frac{\delta F_0[\rho_0]}{\delta\rho_0(\vq)}
%\end{equation}
%in order to introduce all the details of the matrix through a functional
%dependence on $\rho_0(\vq)$.
Notice that, apart from the standard ideal free-energy functional, there
is a non-trivial term arising from the interaction between the fluid and
the matrix.

So far for the intrinsic free-energy functional. Now to obtain the 
functional $F[\rho_1;\mathcal{Q}]$ ---or, considering that we are
describing a QA system, better $F[\rho_1;\rho_0]$--- we make a Legendre transformation
of $\Omega[u_1,u_{10};\mathcal{Q}]$ only w.r.t.\ $u_1$ ($u_{10}$ is assumed
fixed) to obtain
\begin{equation}
\begin{split}
F[\rho_1;\rho_0]=&\,F^{\rm id}[\rho_1]+\frac{T}{T_0}\int\mathrm{d}\vx\,\rho_1(\vx)
\Big\{F_0[\rho_0(\vx,\cdot)]-F_0[\rho_0]\Big\} \\
&+\frac{T}{T_0}\int\mathrm{d}\vx\,\rho_1(\vx)\int\mathrm{d}\vq\,
\left\{\rho_0(\vq)\frac{\delta F_0[\rho_0]}{\delta \rho_0(\vq)}-
\rho_0(\vx,\vq)\frac{\delta F_0[\rho_0(\vx,\cdot)]}{\delta \rho_0(\vx,\vq)}\right\}.
\end{split}
\end{equation}

\subsection{The special case of an ideal matrix}

One particular case which has received some attention in the literature
\cite{given:1992a,lomba:1993,rosinberg:1994} is the case in which the
matrix is also ideal. The reason is that the
ROZ equations for this system can be exactly solved (when $u_{10}$ is
a hard-sphere potential), and, in spite of its simplicity, the blocking
part of the direct correlation function is non-zero.

If the matrix is a configuration of an ideal gas at temperature $T_0$,
then
\begin{equation}
F_0[\rho_0]=kT_0\int\mathrm{d}\vq\,\rho_0(\vq)\Big\{\ln\Big(\mathcal{V}_0
\rho_0(\vq)\Big)-1\Big\}.
\end{equation}
Substituting this $F_0$ in the expressions of the previous section one gets
\begin{equation}
\begin{split}
\rho_0(\vq) &=\frac{e^{\beta_0u_0(\vq)}}{\mathcal{V}_0},\\
\rho_0(\vx,\vq) &=\frac{e^{\beta_0u_0(\vq)+\beta u_{10}(\vx,\vq)}}{\mathcal{V}_0}
=\rho_0(\vq)e^{\beta u_{10}(\vx,\vq)};
\end{split}
\end{equation}
therefore
\begin{eqnarray}
\label{eq:psi_ideal}
F[\rho_1;\rho_0] &=& F^{\rm id}[\rho_1]-kT\int\mathrm{d}\vx
\int\mathrm{d}\vq\,\rho_1(\vx)
\rho_0(\vq)f_{10}(\vx,\vq), \\
f_{10}(\vx,\vq) &\equiv& e^{\beta u_{10}(\vx,\vq)}-1,
\end{eqnarray}
and 
\begin{equation}
\rho_1(\vx)=\frac{e^{\beta u_1(\vx)}}{\mathcal{V}_1}\exp\left\{\int\mathrm{d}\vq\,
\rho_0(\vq)f_{10}(\vx,\vq)\right\}.
\end{equation}

A simple inspection of the exact functional (\ref{eq:psi_ideal}) for an
ideal fluid in an ideal matrix reveals that the non-ideal term is quadratic,
so the only second derivative that is nonzero is $c_{10}=f_{10}$.
Nevertheless, as mentioned above, this system
has a non-trivial $c_\mathrm{b}$ \cite{given:1992a,lomba:1993,rosinberg:1994},
which certainly cannot be derived from (\ref{eq:psi_ideal}) by any functional
differentiation. In spite of this, the functional (\ref{eq:psi_ideal}) contains
all the equilibrium thermodynamics of the system.

\section{Discussion and conclusions}

We have made a first-principles derivation of a density functional formalism for
fluids inside quenched disorder matrices without resorting to the replica trick.
The main conclusion is that, for fixed interaction potential between the
fluid particles, $\mathcal{H}_N(\vx_1,\dotsc,\vx_N)$, and fixed distribution
of the disorder, there exist a unique functional $\mathcal{F}[\rho_1,\rho_{10};\mathcal{Q}]$
from which all the equilibrium structure information and thermodynamics can be derived.  
Given the generalized external potential acting on the fluid particles,
$u_1(\vx)=\mu_1-\varphi_1(\vx)$, and the interaction potential between
the fluid and matrix particles, $u_{10}(\vx,\vq)=-\varphi_{10}(\vx,\vq)$,
the disorder-average of the equilibrium density profile of the fluid,
$\rho^\mathrm{eq}_1(\vx)$, and the fluid-matrix pair distribution
$\rho^\mathrm{eq}_{10}(\vx,\vq)$ can be derived from Eqs.~(\ref{eq:EL}).
Once we have $\rho^\mathrm{eq}_1(\vx)$ and $\rho^\mathrm{eq}_{10}(\vx,\vq)$,
the average of the ``intrinsic'' free-energy of the system is given by
$\mathcal{F}[\rho^\mathrm{eq}_1,\rho^\mathrm{eq}_{10};\mathcal{Q}]$,
and the grand potential by
\begin{equation}
\begin{split}
\overline{\Omega}=&\,\mathcal{F}[\rho^\mathrm{eq}_1,\rho^\mathrm{eq}_{10};\mathcal{Q}]
-\int\mathrm{d}\vx\, \left.\frac{\delta \mathcal{F}[\rho_1,\rho_{10};
\mathcal{Q}]}{\delta \rho_1(\vx)}
\right|_{\substack{\rho_1=\rho^\mathrm{eq}_1\\
\rho_{10}=\rho^\mathrm{eq}_{10}}}\rho^\mathrm{eq}_1(\vx) \\
& -\int\mathrm{d}\vx\mathrm{d}\vq\, \left.\frac{\delta \mathcal{F}[\rho_1,\rho_{10};
\mathcal{Q}]}{\delta
\rho_{10}(\vx,\vq)}\right|_{\substack{\rho_1=\rho^\mathrm{eq}_1\\
\rho_{10}=\rho^\mathrm{eq}_{10}}}\rho^\mathrm{eq}_{10}(\vx,\vq).
\end{split}
\end{equation}

Although for the ideal fluid in a quenched matrix we have been able to derive
the explicit form of $\mathcal{F}[\rho_1,\rho_{10};\mathcal{Q}]$, this is a formidable task
for an arbitrary system. Note that this functional is valid for \emph{any} interaction
potential between the fluid and matrix particles and if we had it, then we would have
solved a very general problem. Thus, it is more practical to turn to a less general
functional, $F[\rho_1;\mathcal{Q}]$, which will be a functional only of $\rho_1(\vx)$
and whose functional form will depend on $u_{10}(\vx,\vq)$.
Again, we have an Euler-Lagrange equation to obtain the equilibrium properties for
a given generalized external potential $u_1(\vx)$ [Eq.~(\ref{eq:dual})], but now
$F[\rho^\mathrm{eq}_1;\mathcal{Q}]$ is not just the average over
disorder of the ``intrinsic'' free energy, but it also contains an additional
contribution due to the quenched-average of the interaction energy between the
fluid and matrix particles, $-\int\mathrm{d}\vx\mathrm{d}\vq\, u_{10}(\vx,\vq)
\rho^\mathrm{eq}_{10}(\vx,\vq)$,
where $\rho^\mathrm{eq}_{10}(\vx,\vq)$ can be obtained from the OZ Eq.~(\ref{ec:OZ10}).

One of the most relevant contributions of this work is the identification
of the direct correlation functionals appearing in the ROZ equations
with second functional derivatives of $F[\rho_1;\mathcal{Q}]$ [Eqs.~(\ref{eq:cc})
and (\ref{eq:c10})]. It is worth mentioning that, in contrast to the case
of classical DFT, the second derivative of $F^\mathrm{ex}[\rho_1;\mathcal{Q}]$
with respect to $\rho_1(\vx)$ and $\rho_1(\vx')$ is not $c_{11}(\vx,\vx';[\rho_1;
\mathcal{Q}])$
but only its connected part. Notwithstanding, the formalism is closed in
the set of correlation functionals $\{(c_{00},h_{00}),(c_{10},h_{10}),(c_{01},h_{01})
,(h_\mathrm{c},c_\mathrm{c})\}$, since the \emph{direct} correlation functionals
are obtained by simple functional differentiations of $F[\rho_1;\mathcal{Q}]$ and
the \emph{total} correlation ones can be derived from the ROZ
Eqs.~(\ref{ec:OZ00}--\ref{ec:OZ01}), and (\ref{ec:OZc}), which have been obtained
as functional identities in the QA-DFT presented in this work.

As we have discussed previously, the blocking parts
$h_\mathrm{b}$ and $c_\mathrm{b}$ do not enter anywhere in the formalism.
The fact that the QA-DFT does not contain these correlations and that
the thermodynamics can be entirely derived from it implies that the 
blocking correlations are not relevant for the thermodynamics. In this
respect, we would like to mention that mode-coupling theory has been recently
extended to QA systems~\cite{krakoviack:2005} in order to study the dynamics
of confined glass-forming liquids, and the only equilibrium structural
information needed to obtain the relaxing density fluctuations
is the set $\{c_\mathrm{c},c_{10},c_0\}$. Thus even the liquid-glass
transition can be determined if we know the functional $F[\rho_1;\mathcal{Q}]$.
%Anyway, if the blocking parts $h_\mathrm{b}$ and $c_\mathrm{b}$ are
%required for some purpose, they can be obtained from Eq.~(\ref{ec:OZ11})
%using the information derived from functional $F[\rho_1]$ and
%one additional closure relation, which is rather simple than solving the
%complete set of ROZ equations.
Also notice that in Refs.~\cite{menon:1994,thalmann:2000}, where freezing
is studied with a Ramakrishnan-Yussouff density functional, the direct
correlation employed in its construction (which is derived with the replica
trick) is $c_\mathrm{c}$, not $c_{11}$.

Finally, as it happens in classical DFT, there are few systems for which
$F[\rho_1;\mathcal{Q}]$ can be obtained exactly (in this case only ideal
fluids in arbitrary matrices, as far as we know).
Therefore, this formalism should be complemented with approximations for
$F^\mathrm{ex}[\rho_1;\mathcal{Q}]$. In this line are the works by
Schmidt and collaborators \cite{schmidt,schmidt:rDFTapp,schmidt:2003},
which make use of the constructing principle of fundamental measure
theory~\cite{rosenfeld:1989,tarazona:1997,tarazona:2000,lafuente:LFMT},
namely the exact result for a 0D cavity (a cavity which can hold at most
either a fluid or a matrix particle) to approximate $F[\rho_1;\mathcal{Q}]$.
Although the results obtained seem promising, we think that the
extension of fundamental measure theory to QA systems involves subtleties
concerning the correlations between fluid and matrix particles that
are difficult to deal with, and further study is required.

Another research line worth exploring is, in analogy to the development of
classical DF approximations, to study the extension of those approximation
based on the thermodynamics and structural information of the
uniform fluid (usually obtained from integral equation theory)
such as the weighted density or the effective liquid
approximations~\cite{evans:1992}. This will be the subject of a forthcoming
work.

\acknowledgments

We appreciate many useful discussions with Matthias Schmidt, Martin-Luc
Rosinberg, Mar\'{\i}a Jos\'e Fernaud and Bob Evans. This work is part of
project BFM2003-0180 from Ministerio de Ciencia y Tecnolog\'{\i}a (Spain),
of project
UC3M-FI-05-007 from Universidad Carlos III de Madrid and Comunidad
Aut\'onoma de Madrid (Spain), and of project MOSSNOHO (S-0505/ESP/000299)
from Comunidad Aut\'onoma de Madrid (Spain).

\end{document}